\documentclass[aps,prl,reprint,superscriptaddress,amsmath,amssymb,showpacs,floatfix]{revtex4-1}
\usepackage{graphicx}
\usepackage{bm}
\usepackage{color}

\begin{document}

\newcommand{\bmq}{{\bm q}}
\newcommand{\bmf}{{\bm f}}
\newcommand{\Wcal}{{\mathcal W}}
\newcommand{\Qcal}{{\mathcal Q}}
\newcommand{\Ecal}{{\mathcal E}}
\newcommand{\Gcal}{{\mathcal G}}
\newcommand{\Fcal}{{\mathcal F}}
\newcommand{\Rcal}{{\mathcal R}}
\newcommand{\Scal}{{\mathcal S}}
\newcommand{\Amat}{{\mathsf A}}
\newcommand{\Fmat}{{\mathsf F}}
\newcommand{\Imat}{{\mathsf I}}

\title{Fluctuation Relation for Heat}

\author{Jae Dong Noh}
\affiliation{Department of Physics, University of Seoul, Seoul 130-743,
Republic of Korea}
\affiliation{School of Physics, Korea Institute for Advanced Study,
Seoul 130-722, Republic of Korea}

\author{Jong-Min Park}
\affiliation{Department of Physics, University of Seoul, Seoul 130-743,
Republic of Korea}

\date{\today}

\begin{abstract}
We present a fluctuation relation for heat dissipation 
in a nonequilibrium system. A nonequilibrium work is known to 
obey the fluctuation theorem in any time interval $t$.
A heat, which differs from a work by an energy change, is shown to
satisfy a modified fluctuation relation. Modification is brought by
correlation between a heat and an energy change during nonequilibrium
processes whose effect may not be negligible even in the $t\to\infty$ limit. 
The fluctuation relation is derived
for overdamped Langevin equation systems, 
and tested in a linear diffusion system.
\end{abstract}
\maketitle

Fluctuations of thermodynamic quantities of 
nonequilibrium systems obey a universal relation referred to as fluctuation 
theorem~(FT)~\cite{Evans93,Lebowitz99,Jarzynski97,Kurchan98,Maes99,
Crooks99,Hatano01,Seifert05,Esposito10}. 
Discovery of the FT leads to a great advance 
in nonequilibrium statistical mechanics. 
Based on the FT, one can generalize the fluctuation dissipation relation 
to nonequilibrium systems~\cite{Harada05,Prost09,Mallick11} and 
figure out fluctuations observed in experimental small-sized
systems~\cite{Carberry04,Collin05,Garnier05}.

The FT for a quantity $\Rcal$ over a time interval $t$ 
takes the form $\langle e^{-\Rcal}\rangle = 1$, where the average $\langle
\cdot \rangle$ is taken over a probability distribution for an initial state
and over all time trajectories. Some quantities further satisfy the FT in the
form $P_r(R)/\tilde{P}_r (-R) = e^R$ where 
$P_r(R) = \langle \delta(\Rcal-R)\rangle$ is a probability density
function~(PDF) for a nonequilibrium process and $\tilde{P}_r(R)$ 
for a corresponding reverse process. The latter is called the detailed
FT and implies the former called the integral FT. 

Consider a system being in thermal equilibrium with a heat
reservoir. We will set the temperature and the Boltzmann constant to unity.
The system is driven into a nonequilibrium state if one adds 
a nonconservative force or applies a time-dependent perturbation. 
Then, there exist 
nonzero net flows of a nonequilibrium work $\Wcal$ into the system and
a heat dissipation $\Qcal$ into the reservoir. It is well established that 
the work $\Wcal$ over a time interval $t$ obeys the 
FT~\cite{Jarzynski97,Crooks99}. In addition, the total
entropy change $\Delta \Scal_{tot} = \Delta \Scal_{sys}+\Delta \Scal_{res}$
with the system~(reservoir) entropy $\Scal_{sys}~(\Scal_{res})$ 
satisfies the integral FT for an arbitrary initial state, and even the
detailed FT for a steady state initial condition~\cite{Seifert05}. 
Thermodynamic quantities are measurable experimentally from time trajectories
in classical systems~\cite{Ciliberto10},
while their experimental measurability in quantum systems is still an open
issue~\cite{Esposito09}.

Fluctuations of heat $\Qcal$, or entropy production 
$\Delta \Scal_{res}=\Qcal/T$, 
has also been attracting much
interest~\cite{Farago02,vanZon03,Visco06,Puglisi06,Baiesi06,Harris06,
Saito07,Rakos08,Fogedby11}. 
Note that a heat differs from a work by an energy change $\Delta \Ecal$ as 
$\Qcal = \Wcal - \Delta \Ecal$. 
When $t$ becomes large, the system will reach a
steady state with constant work and heat production rates on
average. Hence one may expect the FT for heat in the large
$t$ limit where an energy change can be 
negligible~($\Qcal \simeq \Wcal \gg \Delta \Ecal$). In fact, the 
FT for the heat production rate $(\Qcal/t)$ is derived formally 
in the $t\to\infty$ limit~\cite{Kurchan98,Lebowitz99}. 
On the other hand, some model studies demonstrate the FT for
heat~\cite{Saito07} or failure of the FT in the $t\to\infty$ 
limit~\cite{vanZon03,Harris06,Rakos08}. So, it is interesting to
understand how and why the FT is violated for finite $t$
and whether it is restored in the large $t$ limit~\cite{Baiesi06,Puglisi06}.

In this Letter, we present a fluctuation relation for heat, 
given in Eq.~(\ref{FT_Q}). For any process, fluctuations are constrained 
by the energy conservation $\Qcal = \Wcal - \Delta \Ecal$. So a correlation
between thermodynamic quantities plays an important role in characterizing 
the heat fluctuation. We find that the heat distribution satisfies a
modified fluctuation relation that differs from the ordinary FT by a factor
reflecting such a correlation. The fluctuation relation is confirmed for a
linear diffusion system analytically and numerically. Our work provides an
insight into origin for failure of the FT for heat for finite-$t$ interval
and possibly for infinite-$t$ interval. 

We consider a dynamical system described by an overdamped Langevin equation
\begin{equation}\label{e_of_motion}
\frac{d{\bm q}(t)}{dt} = {\bm f}({\bm q}(t)) + {\bm \xi}(t)
\end{equation}
where ${\bm q}=(q_1,q_2,\cdots, q_d)^T$ is a configuration vector, ${\bm
f}({\bm q}) = (f_1({\bm q}),f_2({\bm q}),\cdots,f_d({\bm q}))^T$ is a force, 
and ${\bm \xi} = (\xi_1,\xi_2,\cdots,\xi_d)^T$ is a white noise with
\begin{equation}
\langle \xi_i(t) \rangle = 0 \ , \ \langle \xi_i (t) \xi_j (t') \rangle = 2
\delta_{ij} \delta(t-t')
\end{equation} 
A damping coefficient and a noise strength are set to unity by rescaling
$t$ and ${\bm q}$ properly. The force can be decomposed as $\bmf = \bmf_c +
\bmf_{nc}$, where ${\bm f}_c(\bmq) = -{\bm \nabla}_{\bmq}
\Phi(\bmq)$ is a conservative force with a scalar potential energy function 
$\Phi(\bmq)$ and ${\bm f}_{nc}(\bmq)$ is a nonconservative
force. In this work, we focus on systems with a time-independent potential.
We assume that the system is in thermal equilibrium following the Boltzmann
distribution $P_{eq}({\bm q}) \propto e^{-\Phi({\bm q})}$ initially at
$t=0$. Then it evolves into a nonequilibrium state due to the nonconservative 
force.

When the system follows a path ${\bm q}(\tau)$ for a time interval $0\leq
\tau \leq t$, a nonequilibrium work done by the nonconservative
force, a heat dissipation, and an energy change are given by
functionals 
$\mathcal{W}[{\bm q}(\tau)] = \int_0^t d\tau \dot{\bm q}(\tau) \cdot {\bm
f}_{nc}({\bm q}(\tau))$, 
$\mathcal{Q}[{\bm q}(\tau)] = \int_0^t d\tau \dot{\bm q}(\tau) \cdot {\bm
f}({\bm q}(\tau))$, and $\Delta \Ecal[\bmq(\tau)] =
\Phi(\bmq(t))-\Phi(\bmq(0))$, respectively~\cite{Kwon11}. 
They satisfy the energy conservation 
$\Delta \mathcal{E} = \mathcal{W} - \mathcal{Q}$.
Among these, the PDF for work $P_w(W) \equiv
\langle \delta(\mathcal{W}[{\bm q}(\tau)]-W) \rangle$
satisfies the FT~\cite{Crooks99}
\begin{equation}\label{FT_W}
\frac{P_w(W)}{P_w(-W)} = e^{W} \ .
\end{equation}

Recently, it was found that joint probabilities for thermodynamic quantities
also satisfy similar fluctuation relations~\cite{Garcia10}. Let 
$\{\mathcal{A}_i[\bmq(\tau)]\}$ be a set of functionals 
whose sum is equal to the work 
$\Wcal[\bmq(\tau)] = \sum_i \mathcal{A}_i[\bmq(\tau)]$ and
$\mathcal{A}_i[\bar\bmq(\tau)] = -\mathcal{A}_i[\bmq(\tau)]$ where
$\bar\bmq(\tau) = \bmq(t-\tau)$ is a time-reversed path.
Then, it was found that the joint PDF $P(\{A_i\}) \equiv \langle
\prod_i\delta(\mathcal{A}_i[\bmq]-A_i)\rangle$ satisfies~\cite{Garcia10}
\begin{equation}\label{GFT_general}
\frac{P(\{A_i\})}{P(\{-A_i\})} = e^{\sum_i A_i} \ .
\end{equation}
Those relations reproduce Eq.~(\ref{FT_W}), 
and provide more detailed informations on
nonequilibrium fluctuations~\cite{Garcia12}. 

We apply the formalism to the study of heat fluctuations. 
Consider the joint PDF
$P_{h,e}(Q,\Delta E) \equiv \langle \delta(\Qcal[\bmq]-Q)\delta(\Delta
\Ecal[\bmq]-\Delta E)\rangle$. Since $\Wcal = \Qcal + \Delta \Ecal$, 
it follows from Eq.~(\ref{GFT_general}) that 
\begin{equation}\label{GFT_he}
P_{h,e}(Q,\Delta E) = e^{Q+\Delta E} P_{h,e}(-Q,-\Delta E) \ ,
\end{equation}
which will be referred to as a generalized FT~(GFT). 
One may consider another joint PDF 
$P_{w,e}(W,\Delta E) \equiv \langle
\delta(\Wcal[\bmq]-W)\delta(\Delta\Ecal[\bmq]-\Delta E)\rangle$.
They are related as
\begin{equation}\label{he_to_we}
P_{h,e}(Q,\Delta E) = P_{w,e}(Q + \Delta E,\Delta E) \ .
\end{equation}
So the GFT for $P_{w,e}$ takes a slightly different form as
\begin{equation}\label{GFT_we}
P_{w,e}(W,\Delta E) = e^{W} P_{w,e}(-W,-\Delta E) \ .
\end{equation}

The PDF $P_h(Q) \equiv \langle \delta(\Qcal[\bmq]-Q)\rangle$ for heat
is reduced from $P_{h,e}(Q,\Delta E)$. Integrating both sides of
Eq.~(\ref{GFT_he}) over $(\Delta E)$, we obtain 
a fluctuation relation for the heat:
\begin{equation}\label{FT_Q}
\frac{P_h(Q)}{P_h(-Q)} =  e^{Q} / \Psi(Q)  \ ,
\end{equation}
where
\begin{equation}\label{FT_Q_devi}
\Psi(Q) \equiv 
\int d(\Delta E)\ e^{-\Delta E} P_{e|h}(\Delta E| Q) \ . 
\end{equation}
Note that $P_{e|h}(\Delta E|Q) = P_{h,e}(Q,\Delta E)/P_h(Q)$ 
denotes a conditional probability for an energy change $\Delta E$ to a given
value of heat dissipation $Q$. 
A reciprocity relation $\Psi(-Q) = \Psi(Q)^{-1}$ was used in Eq.~(\ref{FT_Q}). 
The integral version is obtained from Eq.~(\ref{FT_Q})  or
Eq.~(\ref{GFT_he}). 
It is given by 
\begin{equation}\label{IFT_Q}
\left\langle e^{-\Qcal[\bmq(\tau)]} \right\rangle = \left\langle e^{-\Delta
\Ecal[\bmq(\tau)]} \right\rangle \ .
\end{equation}

The detailed FT for heat is modified by the factor $\Psi(Q)$.
The original FT requires that $\Psi(Q)=1$ for all $Q$. 
However one, in general, expects a correlation between $Q$ and
$\Delta E$. Such a correlation leads to a $Q$-dependence
in $\Psi(Q)$, hence invalidates the detailed FT for finite $t$. 

The fluctuation relations can be rewritten in terms of moment generating
functions 
$\Gcal_{\hat w}(\lambda) \equiv \langle e^{-\lambda \Wcal}\rangle$,
$\Gcal_{\hat h}(\eta) \equiv \langle e^{-\eta \Qcal}\rangle$, 
$\Gcal_{\hat{w},\hat{e}}(\lambda,\kappa) \equiv \langle 
e^{-\lambda \Wcal - \kappa \Delta \Ecal}\rangle$, and
$\Gcal_{\hat{h},\hat{e}}(\eta,\kappa) \equiv \langle 
e^{-\eta \Qcal - \kappa \Delta \Ecal}\rangle$. 
All of them are not independent, but are deduced from a single one, 
e.g., $\Gcal_{\hat{w},\hat{e}}$: Equation~(\ref{he_to_we}) yields that 
$\Gcal_{\hat{h},\hat{e}}(\eta,\kappa) = 
\Gcal_{\hat{w},\hat{e}}(\eta,\kappa-\eta)$, and
$\Gcal_{\hat{w}}(\lambda) = \Gcal_{\hat{w},\hat{e}}(\lambda,\kappa=0)$
and $\Gcal_{\hat{h}}(\eta) = \Gcal_{\hat{h},\hat{e}}(\eta,\kappa=0)$. 
The GFTs in Eqs.~(\ref{GFT_he}) and (\ref{GFT_we}) are equivalent to
\begin{eqnarray}
\Gcal_{\hat{w},\hat{e}}(\lambda,\kappa) &=&
\Gcal_{\hat{w},\hat{e}}(1-\lambda,-\kappa) \ , \label{GFT_g} \\
\Gcal_{\hat{h},\hat{e}}(\eta,\kappa) &=&
\Gcal_{\hat{h},\hat{e}}(1-\eta,1-\kappa)  \ .
\end{eqnarray}
Setting $\kappa=0$ in Eq.~(\ref{GFT_g}), one recovers the FT for 
work, $\Gcal_{\hat{w}}(\lambda) = \Gcal_{\hat{w}} (1-\lambda)$.
The fluctuation relation for heat cannot be written in a simple form
with generating functions. Instead, the modification factor 
$\Psi(Q)$ can be written as 
\begin{equation}\label{FT_Q_devi_simple}
\Psi(Q) = \Gcal_{h,\hat{e}}(Q,\kappa=1) / \Gcal_{h,\hat{e}}(Q,\kappa=0)
\end{equation}
with $\Gcal_{h,\hat{e}}(Q,\kappa) \equiv \int d(\Delta E) 
e^{-\kappa (\Delta E)} P_{h,e}(Q,\Delta E)$.

In the $t\to\infty$ limit, the FT is formulated in terms of 
the large deviation function~(LDF)~\cite{Lebowitz99}. 
For the heat distribution, it is defined as
\begin{equation}\label{ldf_q}
e_{h}(q) \equiv \lim_{t\to\infty} -\frac{1}{t}\ln P_h(Q=qt) \ .
\end{equation}
Then, Eq.~(\ref{FT_Q}) yields that 
\begin{equation}\label{FT_Q_ldf}
e_{h}(-q) - e_h(q) =  q + \psi(q)
\end{equation}
where
\begin{equation}
\psi(q) = \lim_{t\to\infty} -\frac{1}{t} \ln \Psi(Q=qt) \ .
\end{equation}
We can further simplify it by introducing a LDF 
\begin{equation}
e_{{h},\hat{e}}(q,\kappa) = \lim_{t\to\infty}-\frac{1}{t}\ln
\Gcal_{{h},\hat{e}}(Q=qt,\kappa) \ , 
\end{equation}
which is obtained from the Legendre transformation
\begin{equation}\label{ehe_lt}
e_{h,\hat{e}}(q,\kappa) = \max_{\eta} \{
e_{\hat{h},\hat{e}}(\eta,\kappa)-q\eta\}
\end{equation} 
of a LDF
$e_{\hat{h},\hat{e}}(\eta,\kappa) = \lim_{t\to\infty} - \frac{1}{t} \ln
\Gcal_{h,e}(\eta,\kappa)$.
Combining these, we obtain that
\begin{equation}\label{psi_q}
\psi(q) = e_{h,\hat{e}}(q,\kappa=1) - e_{h,\hat{e}}(q,\kappa=0)  \ .
\end{equation}
This is a central quantity that determines whether the FT
holds for heat in the $t\to\infty$ limit. 

We apply the formalism to a $d=2$ dimensional linear diffusion system where
the force is given by ${\bm f}({\bm q}) = - \mathsf{F}\cdot {\bm q}$ with
a force matrix 
\begin{equation}\label{Fmat_def}
\mathsf{F} = \left( 
\begin{array}{cc} 1 & \varepsilon \\ -\varepsilon & 1 \end{array} \right) \ .
\end{equation}
This model is a specific case of a general linear diffusion system studied
in Ref.~\cite{Kwon11}, where one can find closed form solutions for
various distribution functions.
The purpose of this study is to confirm the fluctuation relations in 
Eqs.~(\ref{GFT_he}), (\ref{GFT_we}), and (\ref{FT_Q})
explicitly, and to understand the effect of the correlation 
on the fluctuation relation.

The force matrix is decomposed into the symmetric part ${\mathsf F}_s =
({\mathsf F}+{\mathsf F}^T)/2 = {\mathsf I}$ and the anti-symmetric part 
${\mathsf F}_a = ({\mathsf F}-{\mathsf F}^T)/2$. Then, the conservative
force, the energy function, and the nonconservative force are given by
${\bm f}_c({\bm q}) = - {\bm q}$, $\Phi({\bm q}) = {\bm q}^T \cdot {\mathsf
F}_s \cdot {\bm q}/2 =  \frac{1}{2} |{\bm q}|^2$,
and ${\bm f}_{nc} = -{\mathsf F}_a \cdot {\bm q} = \varepsilon (-q_2, q_1)^T$,
respectively. So the model describes a particle trapped in an isotropic 
harmonic potential $\Phi$ and driven by a swirling force $\bmf_{nc}$. 
The parameter $\varepsilon$ represents a strength of the driving force. 

The linear diffusion system was studied in
Ref.~\cite{Kwon11} using a path-integral formalism. We extend the
formalism to obtain the joint probability distributions. 
The algebra is straightforward but rather lengthy. So we present
the explicit expression for the moment generation function
$\Gcal_{\hat{w},\hat{e}}(\lambda,\kappa)$ without derivation. 
Details will be published elsewhere~\cite{Noh12}. 
It is given by
\begin{equation}\label{G_we}
\Gcal_{\hat{w},\hat{e}}(\lambda,\kappa) = \Fcal(\lambda,\kappa) \ ,
\end{equation}
where
\begin{equation}\label{F_def}
\Fcal(x,y) \equiv \frac{e^t}{
\cosh({\Omega(x)}t)+\frac{(1-4y^2)+\Omega(x)^2}{2{\Omega(x)}}
\sinh({\Omega(x)}t)}  
\end{equation}
and 
\begin{equation}\label{Omega_def}
\Omega(x) \equiv \sqrt{1 - 4\varepsilon^2 x(x-1)} \ .
\end{equation}
Note that $\Omega(x) = \Omega{(1-x)}$. 
Hence, $\Gcal_{\hat{w},\hat{e}}(\lambda,\kappa)$ satisfies the GFT 
in Eq.~(\ref{GFT_g}).

\begin{figure}
\includegraphics*[width=\columnwidth]{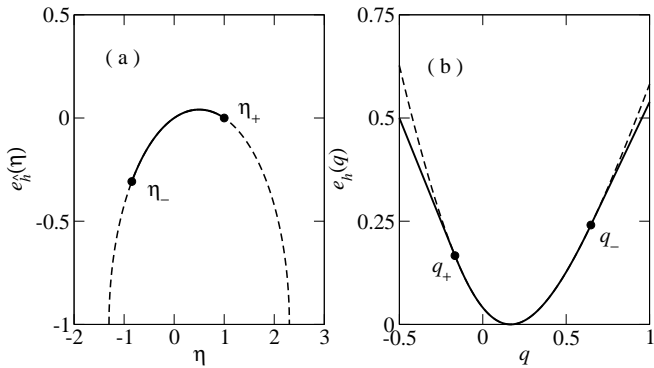}
\caption{Large deviation functions for the generation function in (a) and the
PDF in (b) for heat at $\varepsilon=1/(2\sqrt{3})$ are drawn with solid lines.
Dashed lines are the plot of ($\Omega(\eta)-1$) in (a) and its Legendre
transformation in (b).}
\label{ldf}
\end{figure}

The generating function for $P_h(Q)$ is given by 
\begin{equation}\label{Gh_eta}
\Gcal_{\hat{h}}(\eta) = \Gcal_{\hat{h},\hat{e}}(\eta,\kappa=0) =
\Fcal(\eta,-\eta) \ .
\end{equation}
It does not obey the FT~($\Gcal_{\hat h}(\eta)\neq\Gcal_{\hat h}(1-\eta)$)
for finite $t$.
The corresponding LDF is given by
\begin{equation}\label{ldf_eh}
e_{\hat{h}}(\eta_-\leq \eta\leq \eta_+) 
= \Omega(\eta) - 1 \ .
\end{equation}
We remark that the limit should be taken carefully. 
The function $\Fcal(x,y)$ has a pole singularity at $(\Omega(x)-1)^2 =
4y^2$ in the $t\to\infty$ limit. 
Hence, the LDF is well-defined only within the interval 
$\eta_- \leq \eta \leq \eta_+$ where
$\eta_+=1$ and $\eta_- =
\frac{1}{2}-\frac{1}{2}\sqrt{1+1/\varepsilon^2}$ for $\varepsilon^2>1/3$ and
$\eta_- = 
(\varepsilon^2-1)/(\varepsilon^2+1)$ for $\varepsilon^2\leq 1/3$.
Equation~(\ref{ldf_eh}) is valid only within the interval, 
while $e_{\hat{h}}(\eta)=-\infty$ otherwise. 

The Legendre transformation 
$e_h(q) = \max_\eta \{ e_{\hat{h}}(\eta) - q \eta\}$
yields that
\begin{equation}\label{eh_q}
e_{h}(q) = \left\{
\begin{array}{ll}
-\eta_+ q + \Omega(\eta_+) - 1 & ,\ q \leq q_+ \\ [1mm]
-\eta_- q + \Omega(\eta_-) - 1 & ,\ q \geq q_-  \\ [1mm]
\sqrt{\frac{(1+\varepsilon^2)(q^2+4\varepsilon^2)}{4\varepsilon^2}}
- \frac{q}{2}-1 & , \mbox{ otherwise} 
\end{array}\right.
\end{equation}
where $q_{\pm} = \left. {d e_{\hat h}}/{d\eta}\right|_{\eta=\eta_\pm}$.
The linear branches indicate exponential tails in $P_h(Q)$~\cite{vanZon03}. 

Figure~\ref{ldf}(a) shows the LDF at $\varepsilon=1/(2\sqrt{3})$. 
The function $[\Omega(\eta)-1]$ is drawn with a dashed line, while 
$e_{\hat{h}}(\eta)$ is drawn with a solid line. 
The Legendre transformation $e_h(q)$ is plotted in Fig.~\ref{ldf}(b) 
with a solid line. The Legendre transformation of $[\Omega(\eta)-1]$ is also
drawn with a dashed line. They deviate from each other at $q_{\pm}$.

\begin{figure}
\includegraphics*[width=\columnwidth]{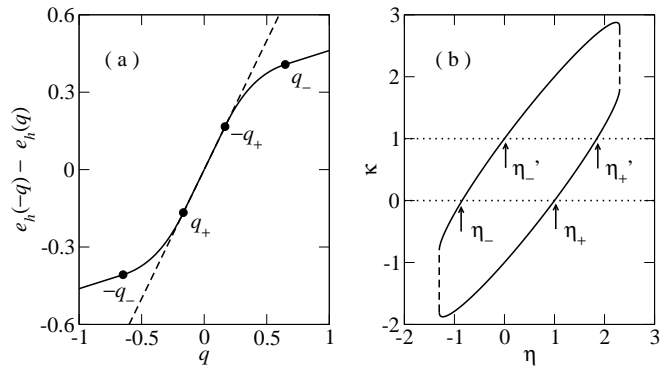}
\caption{(a) Deviation from the FT for heat at $\epsilon=1/(2\sqrt{3})$.
(b) Validity domain for the LDF $e_{\hat{h},\hat{e}}(\eta,\kappa)$. 
}
\label{psi}
\end{figure}

In order to test the FT, we plot $e_h(-q)-e_h(q)$~(solid line) 
in Fig.~\ref{psi}(a). It does not coincide with the dashed straight
line representing $e_h(-q)-e_h(q)=q$ for large $q$, 
which shows that heat does not obey the FT. 
It is worthy to compare our result with that of Ref.~\cite{vanZon03}. 
In both cases, the FT appears to be valid for small values of $q$,
specifically within the domain $|q| \leq |q_+|$ in our study. 
The value of $|e_h(-q)-e_h(q)|$ saturates to a constant for large $|q|$ in 
Ref.~\cite{vanZon03}. It contrasts with the linear increase when $|q|>|q_-|$ 
in our case. It suggests that the heat fluctuations do not exhibit a 
universal behavior~\cite{vanZon03,Rakos08}.

Our theory predicts that $\psi(q)$ describes the
deviation from the FT. We now evaluate $\psi(q)$ explicitly to confirm the
proposed relation in Eq.~(\ref{FT_Q_ldf}). 
Using Eq.~(\ref{he_to_we}) and $\Gcal_{\hat{h},\hat{e}}(\eta,\kappa) =
\Gcal_{\hat{w},\hat{e}}(\eta,\kappa-\eta)$, one has
$\Gcal_{\hat{h},\hat{e}}(\eta,\kappa) = \Fcal(\eta,\kappa-\eta)$. 
So the LDF is given by
\begin{equation}
e_{\hat{h},\hat{e}}(\eta,\kappa) = \Omega(\eta) - 1 \ .
\end{equation}
It appears to be independent of $\kappa$.
However, due to the singularity of $\Fcal$ in Eq.~(\ref{F_def}), 
the LDF is well defined
only within the region $4(\kappa-\eta)^2 \leq (\Omega(\eta)-1)^2$, i.e., 
\begin{equation}
\eta-\frac{1+\Omega(\eta)}{2} \leq \kappa \leq 
\eta + \frac{1+\Omega(\eta)}{2} \ .
\end{equation}
Accordingly, the LDF $e_{\hat{h},\hat{e}}(\eta,\kappa)$ has a $\kappa$ 
dependence. The domain is drawn in Fig.~\ref{psi}(b). 

Now we need perform the Legendre transformation of Eq.~(\ref{ehe_lt}) 
at $\kappa=0$ and $1$. 
When $\kappa=0$, $\eta$ is restricted to the interval
$\eta_- \leq \eta \leq \eta_+$, 
and $e_{h,\hat{e}}(q,\kappa=0)$ becomes equal to
$e_h(q)$ given in Eq.~(\ref{eh_q}). 
When $\kappa=1$, the validity region is shifted to 
$\eta_-'\leq \eta\leq \eta_+'$~(see Fig.~\ref{psi}(b)).  
So, $e_{h,\hat{e}}(q,\kappa=1)$ is given by the
function in Eq.~(\ref{eh_q}) with $\eta_\pm$ and $q_{\pm}$ being replaced
with $\eta_{\pm}'$ and $q_{\pm}'=de_{\hat{h}}/d\eta|_{\eta=\eta_\pm'}$,
respectively. 
Notice the symmetry $\Omega(\eta) = \Omega(1-\eta)$. 
It yields that $\eta_{\pm}' = 1-\eta_{\mp}$ and  
$q_{\pm}' = - q_{\mp}$. Inserting these into Eq.~(\ref{eh_q}), one can find
that $e_{h,\hat{e}}(q,\kappa=1) = e_h(-q)-q$. 
This completes the proof that $\psi(q) = e_h(-q)-e_h(q)-q$.

\begin{figure}
\includegraphics*[width=\columnwidth]{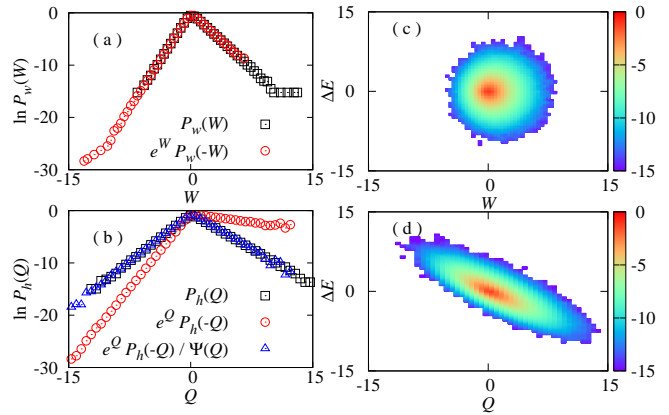}
\caption{(Color online) 
(a) Semi-log plots of $P_w(W)$ and $e^W P_w(-W)$.  
(b) Semi-log plots of $P_h(Q)$, $e^Q P_h(-Q)$, and $e^Q P_h(-Q) / \Psi(Q)$.
Density plots of $\ln P_{w,e}(W,\Delta E)$ in (c) and 
$\ln P_{h,e}(Q,\Delta)$ in (d). }
\label{pdf}
\end{figure}

We also test validity of the relation (\ref{FT_Q}) at finite $t$.
We have solved 
Eq.~(\ref{e_of_motion}) with $\varepsilon=1/\sqrt{3}$ 
numerically $10^7$ times up to $t=0.5$ to measure
various PDFs and $\Psi(Q)$.
In Fig.~\ref{pdf}(a), $P_w(W)$ and $e^W P_w(-W)$ are compared, which
confirms the FT for work. In Fig.~\ref{pdf}(b) 
$P_h(Q)$ displays a disagreement with $e^Q P_h(-Q)$ but
matches perfectly with $e^Q P_h(-Q)/\Psi(Q)$. This is a numerical
verification of the relation in Eq.~(\ref{FT_Q}).
The joint PDF $P_{w,e}(W,\Delta E)$ shown in Fig.~\ref{pdf}(c) is symmetric 
under inversion $\Delta E\to -\Delta E$. 
So the energy may increase or decrease
equally likely irrespective of the amount of work. It explains the reason
why the heat distribution is wider than the work distribution as shown in
Fig.~\ref{pdf}.
One can find an anti-correlation between $\Qcal$ and $\Delta \Ecal$ in
Fig.~\ref{pdf}(d).
Due to the correlation, $\Psi(Q)= \langle e^{-\Delta E}\rangle_Q \neq 1$.

In summary, we have derived the fluctuation relation for heat 
in Eq.~(\ref{FT_Q}) using the GFT in Eq.~(\ref{GFT_he})
for the joint PDF.  The heat distribution does not obey the same type of
the fluctuation relation as the work distribution does.
The modification is given by the factor 
$\Psi(Q)\equiv \langle e^{-\Delta E}\rangle_Q$ that depends on the
correlation between heat and energy change. 
The modified fluctuation relation for the heat has been 
tested analytically and numerically for a linear diffusion system.

Our result shows that the FT for heat is not valid in general 
for finite $t$. The model studies in this work and in Ref.~\cite{vanZon03} 
show explicitly that the FT is violated even for the LDF 
in the $t\to\infty$ limit.
Nevertheless, it still remains as an open question whether there is a
criterion for the FT in terms of the LDF.
A sufficient condition is readily obtained from our result. Suppose that the 
energy function is strictly bounded as $E_0<\Phi(\bmq)<E_1$ with
finite $E_{0,1}$~\cite{Maes99,Kurchan98,Harris06}. 
Then, $e^{-(E_{1}-E_{0})}< \Psi(Q)
<e^{E_{1}-E_{0}}$, hence $\psi(q)=0$ and the FT holds.
Hopefully, our formalism may yield a more strict condition for the FT.
Future works are necessary in order to understand implication of the
proposed fluctuation relation and to generalize it for systems 
with a time-dependent perturbation or systems in contact with many reservoirs.
Experimental studies in small-sized systems~\cite{Ciliberto10} 
are also necessary in order to characterize nonequilibrium fluctuations of heat.

This work was supported by Mid-career Researcher Program through NRF Grant
No.~2011-0017982 funded by the Ministry of Education, Science, and
Technology of Korea. We thank Hyunggyu Park and Chulan Kwon for stimulating
discussions.

\end{document}